# Optimum PID Control of Multi-wing Attractors in A Family of Lorenz-like Chaotic Systems

Anish Acharya[1], Saptarshi Das[2]

1. Department of Instrumentation and Electronics Engineering, Jadavpur University, Salt-Lake Campus, LB-8, Sector 3, Kolkata-700098, India.
2. Department of Power Engineering, Jadavpur University, Salt-Lake Campus, LB-8, Sector 3, Kolkata-700098, India.
Email: saptarshi@pe.jusl.ac.in

Indranil Pan[2,3]

3. MERG, Energy, Environment, Modelling and Minerals ($E^2M^2$) Research Section, Department of Earth Science and Engineering, Imperial College London, Exhibition Road, London SW7 2AZ, UK.

*Abstract*—Multi-wing chaotic attractors are highly complex nonlinear dynamical systems with higher number of index-2 equilibrium points. Due to the presence of several equilibrium points, randomness of the state time series for these multi-wing chaotic systems is higher than that of the conventional double wing chaotic attractors. A real coded Genetic Algorithm (GA) based global optimization framework has been presented in this paper, to design optimum PID controllers so as to control the state trajectories of three different multi-wing Lorenz like chaotic systems viz. Lu, Rucklidge and Sprott-1 system.

## I. Introduction

Chaos is a field in mathematics which has found wide application around us. Chaos theory studies the behavior of dynamical systems which are nonlinear, highly initial condition sensitive, having deterministic (rather than probabilistic) underlying rules which every future state of the system must follow. Such systems exhibit aperiodic oscillations in the time series of state variables. It has a large or infinite number of unstable periodic patterns which is commonly termed as order in disorder. Long term prediction is almost impossible due to the sensitive dependence on initial conditions. Though such effect may seem quite unusual but it is however observed in very simple systems, for example, a ball placed at the crest of a hill might roll into different valleys depending on slight difference in the initial position. Most common chaotic phenomenon is observed in case of regular weather prediction. Other application of chaos theory is pervaded in many fields like geology, mathematics, biology, microbiology, computer science, economics, philosophy, politics, population dynamics, psychology, robotics etc. Some real world applications of chaotic time series are computer networks, data encryption, information processing, pattern recognition, economic forecasting, market prediction etc [1].

Chaotic systems may cause trouble due to their unusual, unpredictable behavior. Hence chaotic control is gaining increasing attention in last few years [1]. In chaotic control, the prime objective is to suppress the chaotic oscillations completely or reduce them to regular oscillations. Nowadays many control techniques such as open loop control methods, adaptive control methods, traditional linear and non linear control methods, fuzzy control techniques etc. are used to control chaotic systems [2]. Chaos control uses the fact that any chaotic attractor which contains infinite number of unstable periodic orbits can be modified using external control action to produce a stable periodic orbit. The chaotic system's states never remains in any of this unstable orbits for long time rather it continuously switches from one orbit to the other which gives rise to this unpredictable, random wandering of the state variables over longer period of time. Chaotic control is basically the stabilization, by means of small system perturbations, of one of these unstable periodic orbits. The result is to render an otherwise chaotic motion more stable and predictable, which is often an advantage. The perturbation must be tiny, to avoid significant modification of the system's natural dynamics. Several techniques have been used to chaos control, but most are developments of two basic approaches: the OGY (Ott, Grebogi and Yorke) method [3], and Pyragas continuous control method [4]. Both methods require a previous determination of the unstable periodic orbits of the chaotic system before the controlling algorithm can be designed. The basic difference between the OGY and Pyragas methods of chaos control is that the former relies on the linearization of the Poincare map and the later is based on time delay feedback. Though PID type controller design has been found in recent literatures like [5] for state synchronization of chaotic systems for different initial condition, but optimum PID control of chaotic systems [6] is not well addressed yet, especially for the control of highly complex chaotic systems like multi-wing attractors in the Lorenz family as attempted in this paper.

Rest of the paper is organized as follows. Section II reports Lorenz family of multi-wing chaotic systems. Section III presents simulation results with GA based optimum PID controller to suppress chaotic oscillations in multi-wing attractors with robustness study in Section IV. The paper ends with conclusion as section V, followed by the references.

## II. Basics of the Multi-wing Chaotic Attractors

### A. Lorenz Family of Multi-wing Chaotic Systems

Three classical examples of symmetric double-wing chaotic attractors are studied here among the Lorenz family of systems.

State equations of the Lorenz family of chaotic systems contain either square and/or cross-terms which can be replaced by a multi-segment parameter adjustable quadratic function (1) to form generate multi-wing attractor with additional flexibility of modifying the number and location of index-2 equilibrium points. As reported in the pioneering work [7] that the segment characteristics like slope and width can be adjusted using the parameters $\{F_0, F_i, E_i\}$ of equation (1). This typical function increases the number of index-2 equilibrium points of Lorenz family of chaotic systems from 2 to $(2N+2)$, thereby increasing the randomness of the state trajectories of nominal (double-wing) chaotic system which is hard to control.

$$f(x) = F_0 x^2 - \sum_{i=1}^{N} F_i \left[ 1 + 0.5 \operatorname{sgn}(x - E_i) - 0.5 \operatorname{sgn}(x + E_i) \right] \quad (1)$$

where, $\operatorname{sgn}(x) = \begin{cases} 1 & \text{for } x > 0 \\ 0 & \text{for } x = 0 \\ -1 & \text{for } x < 0 \end{cases}$ (2)

### B. Chaotic Multi-wing Lu System

The double-wing Lu system [8] is represented by

$$\begin{aligned} \dot{x} &= -ax + ay \\ \dot{y} &= cy - xz \\ \dot{z} &= xy - bz \end{aligned} \quad (3)$$

The typical parameter settings for chaotic double-wing Lu attractor is given by $a = 36, b = 3, c = 20$. The equilibrium points of the Lu system are located at $(0,0,0); (\pm\sqrt{bc}, \pm\sqrt{bc}, c)$. The state equations of the multi-wing chaotic Lu attractor whose states are to be controlled are given by:

$$\begin{aligned} \dot{x} &= -ax + ay \\ \dot{y} &= cy - (1/P) xz + u \\ \dot{z} &= f(x) - bz \end{aligned} \quad (4)$$

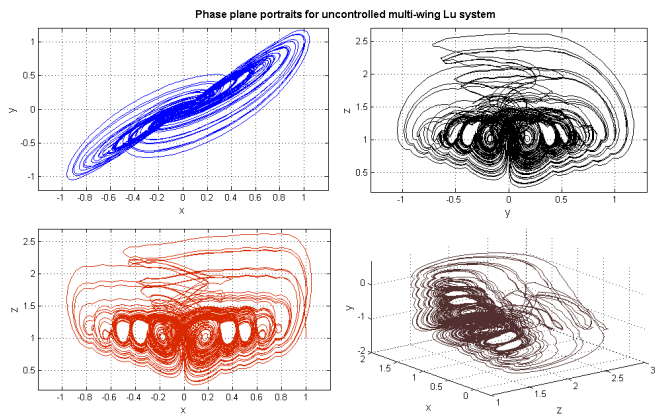

Figure 1. Uncontrolled phase plane portraits for multi-wing Lu system.

Here, $P$ reduces the dynamic range of the attractors so as to facilitate hardware realization. The suggested parameters for $N = 4$ are given below [7] for which the chaotic Lu system exhibits multi-wing attractors in the phase portraits (Fig. 1).
$P = 0.05, F_0 = 100, F_1 = 10, F_2 = 12, F_3 = 16.67, F_4 = 18.18,$
$E_1 = 0.3, E_2 = 0.45, E_3 = 0.6, E_4 = 0.75$

In (4) the PID control action is added to the second state to suppress the chaotic oscillations and is given by (5).

$$u = K_p e + K_i \int e.dt + K_d \frac{de}{dt} \quad (5)$$
$$e = |r - y|$$

Here, $\{K_p, K_i, K_d\}$ are the controller gains which are to be found out by a suitable optimization technique for the reference signal ($r$) as the unit step.

### C. Chaotic Multi-wing Rucklidge system

The double-wing Shimizu-Morioka system [9] is given by

$$\begin{aligned} \dot{x} &= -ax + by - yz \\ \dot{y} &= x \\ \dot{z} &= y^2 - z \end{aligned} \quad (6)$$

The typical parameter settings for chaotic double-wing Shimizu-Morioka attractor is given by $a = 2, b = 7.7$. The equilibrium points of the Shimizu-Morioka system are located at $(0,0,0); (0, \pm\sqrt{b}, b)$. The state equations of the multi-wing chaotic Shimizu-Morioka attractor whose states are to be controlled are given by:

$$\begin{aligned} \dot{x} &= -ax + ay \\ \dot{y} &= (c-a)x + cy - (1/P) xz + u \\ \dot{z} &= f(x) - bz \end{aligned} \quad (7)$$

The above mentioned multi-wing chaotic system is also known as modified Rucklidge system [7]. The suggested parameters for $N = 3$ are given below [7] for which the chaotic Rucklidge system exhibits multi-wing attractors in phase portraits (Fig. 2).

$P = 0.5, F_0 = 4, F_1 = 9.23, F_2 = 12, F_3 = 18.18, E_1 = 1.5, E_2 = 2.25, E_3 = 3.0$

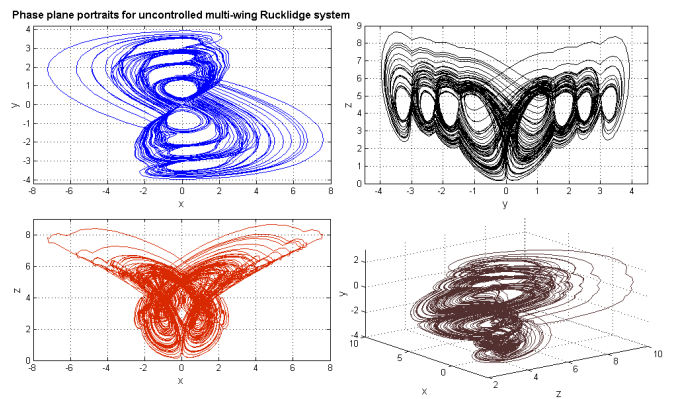

Figure 2. Uncontrolled phase plane portraits for multi-wing Rucklidge (Shimizu-Morioka) system.

## D. Chaotic Multi-wing Sprott-1 system

The double-wing Sprott-1 system [10] is given by

$$\dot{x} = yz$$
$$\dot{y} = x - y \quad (8)$$
$$\dot{z} = 1 - x^2$$

The equilibrium points of the Sprott-1 system are located at $(\pm 1, \pm 1, 0)$. State equations of the multi-wing chaotic Sprott-1 attractor whose states are to be controlled are given by:

$$\dot{x} = yz$$
$$\dot{y} = x - y + u \quad (9)$$
$$\dot{z} = 1 - f(x)$$

The suggested parameters for $N = 4$ are given below [7] for which the chaotic Sprott-1 system exhibits multi-wing attractors in phase portraits (Fig. 3).

$F_0 = 1, F_1 = 5, F_2 = 5, F_3 = 6.67, F_4 = 8.89, E_1 = 2, E_2 = 3, E_3 = 4, E_4 = 5$

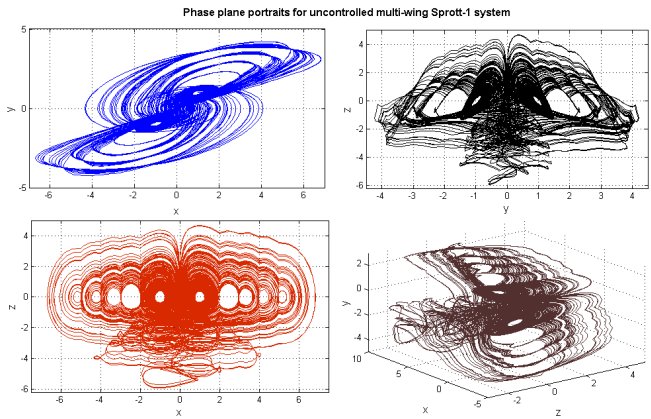

Figure 3. Uncontrolled phase plane portraits for multi-wing Sprott-1 system.

## III. SIMULATION AND RESULTS

Each of the above three multi-wing chaotic systems are to be controlled using a PID controller (5) which will enforce the second state variable ($y$) to track the unit reference step signal ($r$). Instead of simple error minimization criteria for PID controller tuning the well known Integral of Time multiplied Absolute Error (ITAE) has been taken as the performance index ($J$) so as to ensure fast tracking of the second state.

$$J = \int_0^\infty t|e(t)|dt = \int_0^\infty t|r(t) - y(t)|dt \quad (10)$$

For time domain simulation purposes, the upper time limit of the above integral is restricted to realistic values depending on the speed of the chaotic time series to ensure that all oscillations in the state variables have died down due to introduction of the PID control action. It is also seen that controlling the second state variable with PID automatically damps chaotic oscillations in the other two state variables.

Tuning of the PID controller gains have been done in this study using the widely used population based optimizer known as Genetic Algorithm. Due to randomness of the chaotic time series of the multi-wing attractors, the error signal with respect to step command input also becomes highly jittery and will contain several minima which justify the application of GA in such controller tuning problems. For the control of systems, governed by nonlinear differential equations, a GA based PID controller design with other time domain performance index optimization based methods could also have been used like that presented by Das *et al.* [11] but for simplicity we restricted the study with ITAE based PID design only to handle multi-wing attractors in chaotic nonlinear dynamical systems. The real coded GA based optimization (parameters adopted from [11]) results for the PID controller parameters (gains) are given in Table I for the three respective multi-wing attractors among the Lorenz family of chaotic systems.

TABLE I. GA BASED OPTIMUM PID CONTROLLER SETTINGS FOR CHAOS SUPPRESSION IN MULTI-WING ATTRACTORS

| Multi-wing Chaotic systems | $J_{min}$ | $K_p$ | $K_i$ | $K_d$ |
|---|---|---|---|---|
| Lu system | 244.9856 | 3.155719 | 27.56161 | 1.449423 |
| Rucklidge system | 1.160501 | 19.43469 | 30.09687 | 0.237435 |
| Sprott-1 system | 1468.193 | 0.271562 | 0.432607 | 0.393097 |

## A. PID Control of Multi-wing Lu System

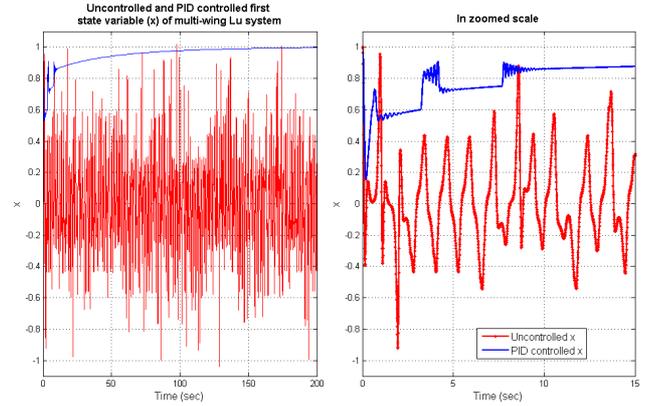

Figure 4. Controlled response of first state variable (x).

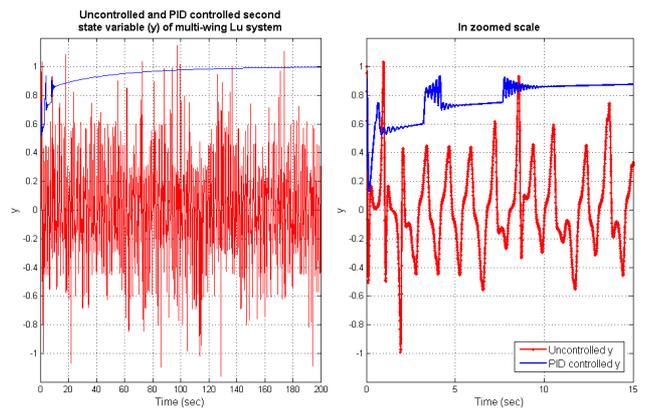

Figure 5. Controlled response of second state variable (y).

Simulation results for the uncontrolled and PID controlled state variables of the multi-wing Lu system (4) has been shown in Fig. 4-6 with the corresponding control signal and error of the second state depicted in Fig. 7. The PID controlled phase portraits in Fig. 8 shows that the presented technique successfully damps wandering of the states which also evident from the individual controlled state trajectories (Fig. 4-6).

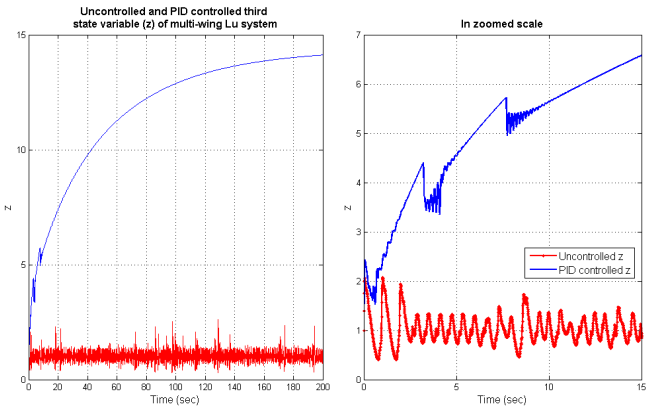

Figure 6. Controlled response of third state variable (z).

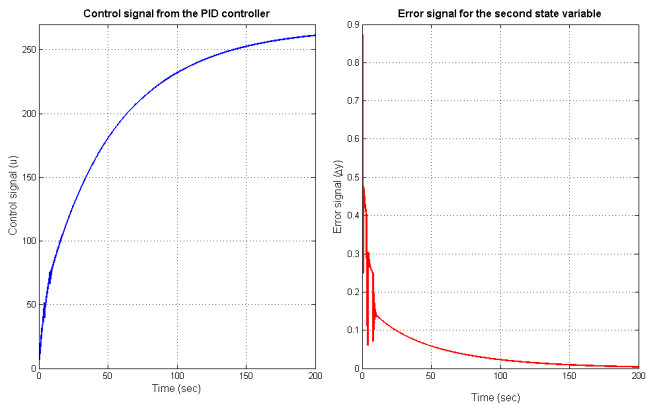

Figure 7. Control and error signal in PID controlled multi-wing Lu system.

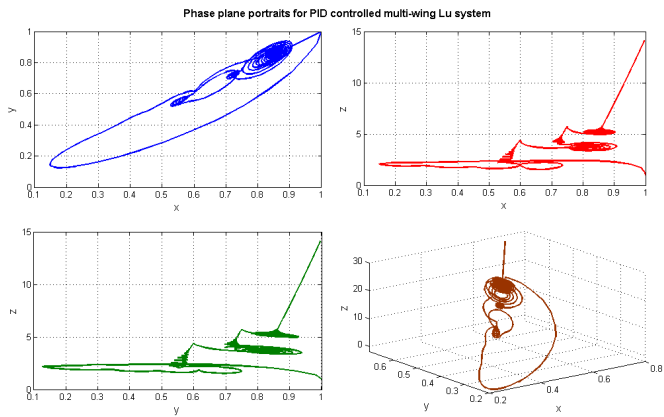

Figure 8. PID controlled phase plane portraits for multi-wing Lu System.

## B. PID Control of Multi-wing Rucklidge System

Similar nature of chaos control can be found the multi-wing Rucklidge system (7) also with the GA based optimum PID controllers which enforces fast tracking of the second state variable. Also the irregular oscillations of this system are found to be more sluggish compared to the Lu system which is controlled by the PID to track a reference using ITAE criteria. Also, controlled state trajectories are smooth at initial stages unlike that for the Lu system. Here, Fig. 9-11 show the state trajectories with the control/error in Fig. 12 and the controlled phase portraits depicted in Fig. 13.

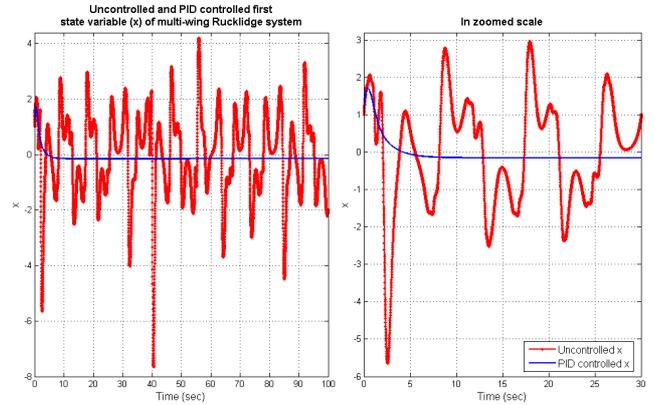

Figure 9. Controlled response of first state variable (x).

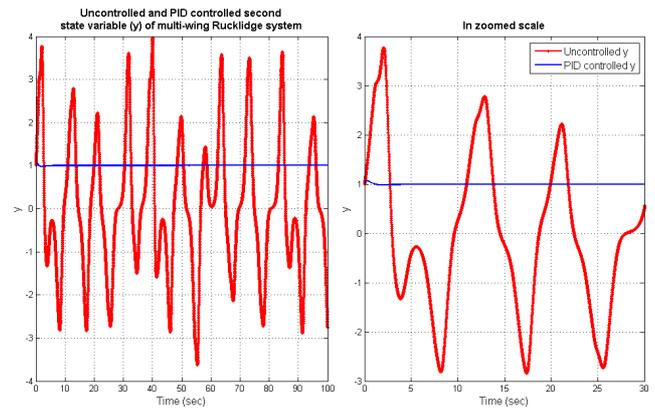

Figure 10. Controlled response of second state variable (y).

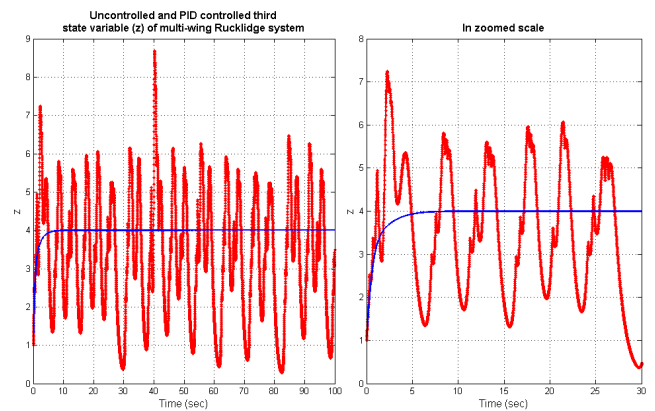

Figure 11. Controlled response of third state variable (z).

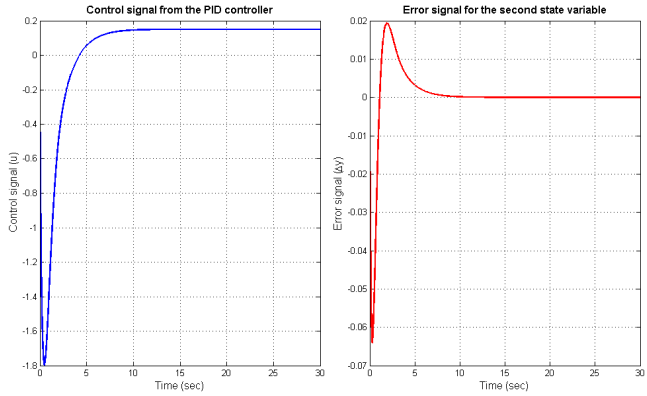

Figure 12. Control & error signal in controlled multi-wing Rucklidge system

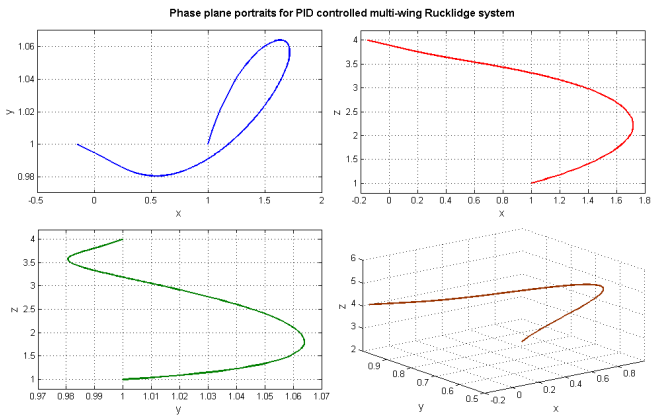

Figure 13. PID controlled phase portraits for multi-wing Rucklidge System

## C. PID Control of Multi-wing Sprott-1 System

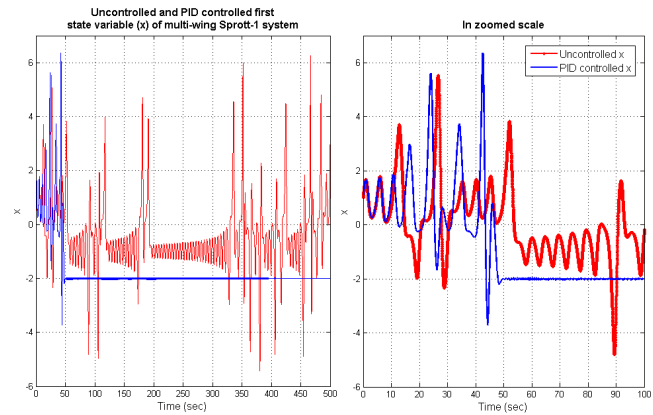

Figure 14. Controlled response of first state variable (x).

For the multi-wing Sprott-1 system (9), the states are even more sluggish where the ITAE based GA tuned PID enforces fast reference tracking and simultaneously damping chaotic oscillations (Fig. 14-16) in an efficient way as can also be seen from the control and error signals in Fig. 17. Wandering of the states can only be found at the initial stages of the phase portraits (Fig. 18), like that in the multi-wing Lu system also. It is well known that chaotic systems are highly sensitive to the initial conditions of the states and in the presented approach only a single value of the states are assumed to tuned the PID controllers. Hence, robustness of the present PID control scheme is shown in next section for respective cases.

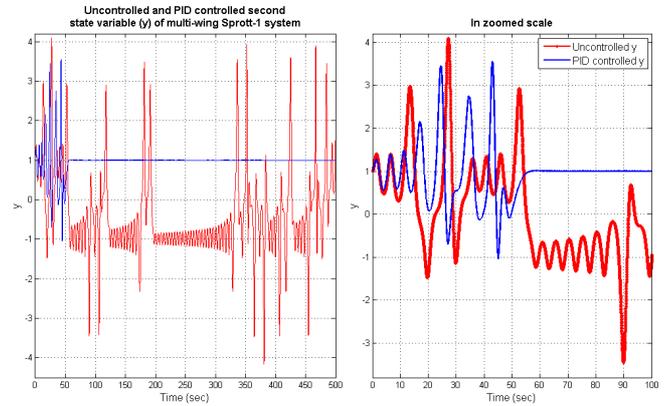

Figure 15. Controlled response of second state variable (y).

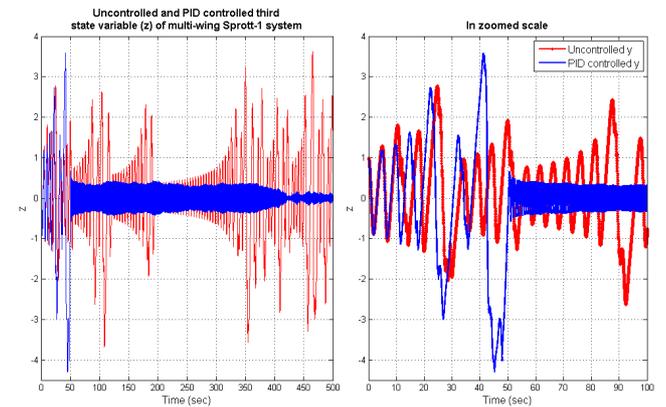

Figure 16. Controlled response of third state variable (z).

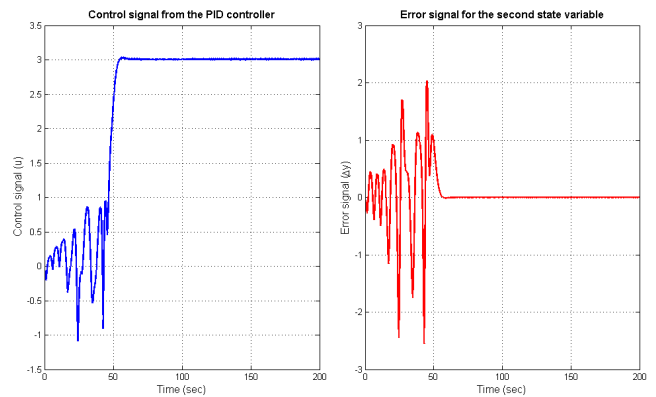

Figure 17. Control and error signal in controlled multi-wing Sprott-1 system

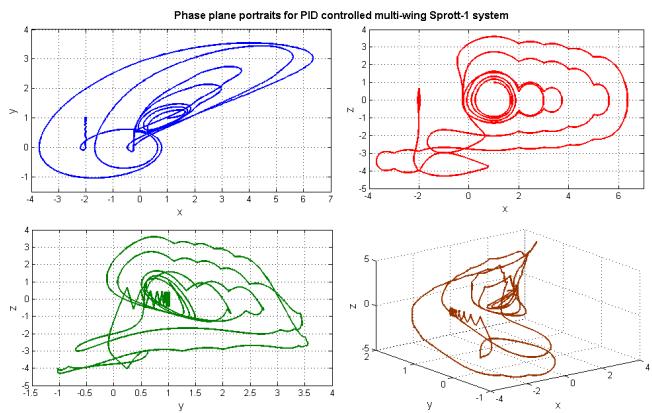

Figure 18. PID controlled phase portraits for multi-wing Sprott-1 System.

## IV. ROBUSTNESS OF THE PID CONTROL SCHEME FOR DIFFERENT INITIAL CONDITIONS OF CHAOTIC ATTRACTORS

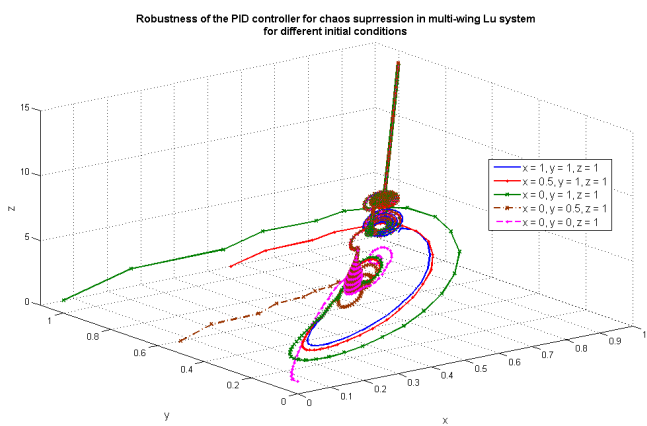

Figure 19. Robustness of PID for controlling multi-wing Lu system.

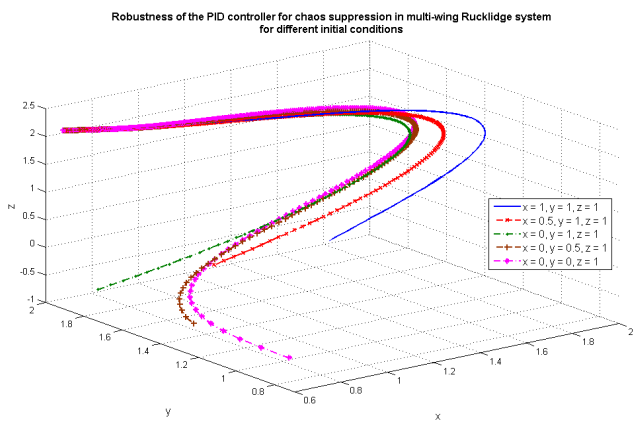

Figure 20. Robustness of PID for controlling multi-wing Rucklidge system.

The proposed PID control scheme has also been found to be robust enough with variation in the initial conditions of multi-wing chaotic systems. Fig. 19-21 shows that in the phase-plane portraits the chaotic oscillations get suppressed along the same trajectory for the three systems, even if the initial conditions for the first two states are gradually decreased from unity to zero.

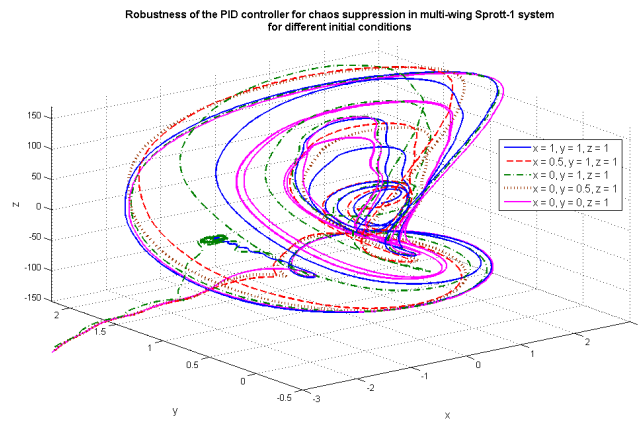

Figure 21. Robustness of PID for controlling multi-wing Sprott-1 system.

## V. CONCLUSION

GA based optimum PID controllers are designed to suppress chaotic oscillations in few highly complex multi-wing Lorenz like chaotic systems. The controller enforces fast tracking of the second state which also damps chaotic oscillation in the other states and found to be robust enough for different initial conditions for such typical nonlinear dynamical systems.